\documentclass[aps,pre,twocolumn,showpacs]{revtex4}

\usepackage{epsfig}

\usepackage{times}
\begin{document}

\title{Vertex dynamics during domain growth in three-state models}
\author{Attila Szolnoki and Gy\"orgy Szab\'o}
\affiliation
{Research Institute for Technical Physics and Materials Science,
P.O. Box 49, H-1525 Budapest, Hungary}

\date{\today}

\begin{abstract}

Topological aspects of interfaces are studied by comparing quantitatively
the evolving three-color patterns in three different models, such as
the three-state voter, Potts and extended voter models. The statistical 
analysis of some geometrical features allows us to explore the role of
different elementary processes during distinct coarsening phenomena in
the above models.
\end{abstract}


\maketitle

The dynamics of ordering process from a disordered state is a
long-standing problem with wide range of application
\cite{gunton:83,bray:adp94}. In many cases the growing domains
can be characterized by a typical length $r(t)$ (average linear
size, correlation length, etc.) for the late stage of coarsening
and on the typical length scales the domain structures become
similar. The time dependence of the linear length scale can be
described by an algebraic growth law, $r(t) \sim t^n$ where
the growth exponent
$n={1 \over 2}$ for the curvature-driven growth if the order 
parameter is not conserved during the elementary processes
\cite{allen:amet79,bray:adp94,brown:pre02}.

During the domain growth the interfaces form closed loops in the
two-state systems \cite{kadanoff:76}. In the $Q$-state ($Q \ge 3$)
systems, however, one can observe vertices where three (or more) states
(or interfaces) meet. According to an early conjecture of Lifshitz
\cite{lifshitz:jetp62} and Safran \cite{safran:prl81} the curvature 
driving force for
such a domain growth is practically switched off along the straight-line
interfaces connecting two vertices and
its absence may affect the dynamics of the domain growth. The
subsequent numerical investigations of the two-dimensional $Q$-state
Potts models did not confirmed this conjecture. 
More precisely, the first Monte Carlo simulations reported $Q$-dependent
effective exponent \cite{sahni:prl83,sahni:prb83},
however, the large scale simulation 
suggested that $n=1/2$ holds independent of the ordering degeneracy, $Q$
\cite{grest:prb88}.
Nevertheless, the numerical evidences to draw any solid conclusions are far
from satisfactory. 
Recently the effects of the branching interfaces
during the coarsening process has also been investigated by 
Cardy \cite{cardy:npb00} using a field theoretical approach.

Besides, a distinct universality class, represented by the
two-dimensional voter model, is introduced to consider the coarsening
process driven by interfacial noises \cite{dornic:prl01}.
In this case the kinetics of domain growth shows a logarithmic
decay of the density of interfaces \cite{frachebourg:pre96,bennaim:pre96}. 
According to this argument this class is identified by the absence of
surface tension. 

Motivated by the above mentioned topological aspect, we will consider numerically
the time dependence of coarsening process for the three-color
growing domain structures. For this purpose we have adopted a
numerical technique developed previously to investigate the geometry
of spiral structures appearing in some cyclically dominated three-state
voter models \cite{ss:pre02}.

In this Brief Report we compare the variation of topological features
during the coarsening dynamics in different three-state models. The investigated
models are the Potts model, the voter model, and a voter model extended by 
Potts energy  \cite{ss:pre02}. 
Common feature of these
models is the existence of three (equivalent) types of growing domains 
separated by
branching interfaces. The comprehensive comparative geometrical analysis
of the  topological features allows us to extend the statistical analyses
yielding a deeper insight into the kinetics of coarsening.

Henceforth we consider a square lattice (with $L \times L$ sites under
periodic boundary conditions) where at each site $x$ 
there is a state variable with three possible states, namely 
$s_x =0$, $1$, or $2$. The time dependence of these state
variables is governed by random sequential updates. Starting from a
random initial state this elementary process is repeated.
The symmetric elementary rules conserve the equivalence between the
three states.

In the case of voter model we choose randomly a nearest neighbor pair
of sites and one of these state variables is changed. 
Variation can
only occur if the randomly chosen states are different. More precisely,
the different state variables $(s_1,s_2)$ become uniform yielding
$(s_1,s_1)$ or $(s_2,s_2)$ pair with equal ($1/2$) probabilities. 
On the two dimensional lattices this model exhibits coarsening with a 
logarithmic decay of density of interfaces
\cite{frachebourg:pre96,bennaim:pre96}.

The energy in the three-state ferromagnetic Potts model is defined
by the Hamiltonian as
\begin{equation}
H = \sum_{<x,y>} [1- \delta (s_x,s_y)]
\label{PottsH}
\end{equation}
where the summation runs over the nearest neighbor sites and 
$\delta(s_x,s_y)$ indicates the Kronecker's delta. On the analogy
of the Glauber dynamics, here the state $s_x$ at a randomly 
chosen site $x$ is updated to a new randomly chosen state
$s^{\prime}_x$. The transition probability is given by 
\begin{equation}
W(s_x \rightarrow s^{\prime}_x) = {1 \over 1 + \exp{(\delta H / T)}}
\label{eq:w}
\end{equation}
where $\delta H$ is the energy difference between the final and 
initial state, and $T$ is temperature. 
The Boltzmann constant is chosen to be unity, as usual. 
This system becomes ordered if $T < T_c=0.995$ \cite{binder:92}.
Our simulations are performed well below the critical point
($T=0.6T_c$) where the 
interfaces are smooth enough to apply geometrical analysis.
At the same time this temperature is sufficiently high
to avoid temporal pinning \cite{fogedby:prb88,castan:prb90} 
or the observation of artifact domain shape as a consequence of 
the specified host lattice.
In all the present models the Potts energy expresses the total length
of interfaces (separating the homogeneous domains) measured in the
unit of lattice constant chosen to be one. If the time dependence of
the excess energy per sites [$\Delta E(t)$] is measured from the
corresponding thermal average value $E_T=\langle H \rangle_T / L^2 
= E(t \to \infty)$, that is 
\begin{equation}
\Delta E(t) = E(t) -E_T \; ,
\label{eq:ex}
\end{equation}
then its inverse [$1 / \Delta E(t)$] estimates
the average domain radius \cite{mouritsen:impb90}.
According to the Allan-Cahn growth law the inverse of $\Delta E(t)$ 
shows an algebraic decay with an exponent of $0.5$.

A very relevant difference between the above mentioned two models is
the presence (absence) of bulk fluctuation in the Potts (voter) model.
The third model is considered as a combination of the standard voter
and Potts models where the adoption of the nearest-neighbor's
opinion (state) is affected by their neighborhood via the Potts energy
\cite{ss:pre02}. This means that the new possible state $s^{\prime}$
for a randomly chosen site $x$ should be equivalent to one of the 
neighboring states (as it happens for the voter model) meanwhile 
the transition probability is defined by the expression (\ref{eq:w}).
Due to this modification domain growth appears for arbitrary 
$T$, while the interfacial irregularities are reduced by the 
surface tension (Potts energy) whose strength is tuned by $T$.
Evidently, the behavior of the standard voter model can be
reproduced by this version in the limit $T \to \infty$.
Henceforth the analysis of this model will be restricted to a fixed
temperature $T=2$. It will be demonstrated that the consequences
of the surface tension can be well observed during the domain
growth for such a high temperature whose value exceeds substantially
the critical temperature of the corresponding Potts model.

\begin{figure}
\centerline{\epsfig{file=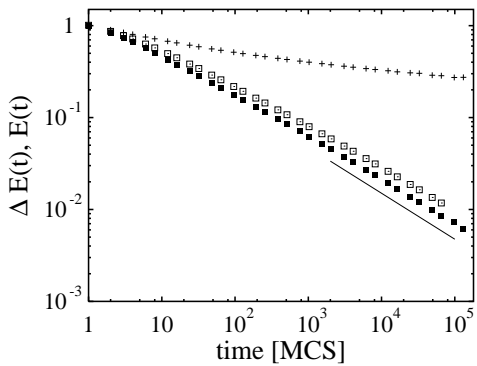,width=6.5cm}}
\caption{\label{fig:f1}The decay of interfacial energy per site $E(t)$ for
the voter (pluses) and extended voter (open boxes) models. For the Potts
model the excess interfacial energy per site [$\Delta E(t)$] is shown
by filled boxes as a function of time. The solid line
indicates the slope of $-0.5$.}
\end{figure}

Our Monte Carlo (MC) simulations were performed for $L=2000$ and the
results were averaged over 20-100 independent runs. For such
a large system size the domain growth could be monitored until
$10^5$ MCS (MC steps per sites) without the disturbance of the finite
size effects. First we consider $E(t)$ (the concentration of domain
walls for the voter and extended voter models) and the excess energy
per site $\Delta E(t)$ (for the Potts model). The comparison of these
quantities on a log-log plot (see Fig.~\ref{fig:f1}) demonstrates
that the logarithmically slow coarsening dynamics of the voter model
can be well distinguished from those situations where the growth
process is affected by the surface tension. Both the $E(t)$ (for the
extended voter model) and the $\Delta E(t)$ (for the Potts model)
tends towards the prediction of algebraic growth law with Allen-Cahn
exponent ($1/t^{1/2}$). 
It demonstrates that the asymptotic behavior can only be seen for
$t > 10^3$ MCS.
Now, it is worth mentioning that our numerical data are consistent
with the appearance of a logarithmic correction ($\ln t / t^{1/2}$)
for $t < 2 \cdot 10^4$ MCS within our statistical error. Unfortunately,
the large statistical error in the last time decade does not allow us
to  extrapolate this behavior for longer times. 
 
\begin{figure}
\centerline{\epsfig{file=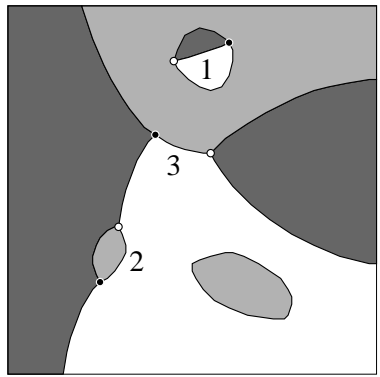,width=5cm}}
\caption{\label{fig:f2}Schematic plot of domain walls of three-state 
topology showing the three possible vertex types. Black (white) bullets
represent vertices (antivertices). The inserted figures denote the number
of different antivertices connected to a given vertex.}
\end{figure}

In order to have a picture about the essence of our topological analysis
Fig.~\ref{fig:f2} illustrates schematically a typical part of the
three-color maps on a continuous background if the motion of interfaces is
characterized by infinitesimally small steps. In this case we can neglect
those vertices were more than three states meet. In such a map the typical
objects are the islands and the three-edge vertices. An (isolated)
island is surrounded by the same domain therefore its boundary is free
of vertices. In fact, two types of vertices (called vertices and
antivertices) can be distinguished depending on whether we find 0, 1, 2
or reversed order of states when going around the center in clock-wise
direction. The vertices and antivertices are located  alternately along
the boundaries. Each vertex can be connected to one, two, or three
antivertices thus they can be further classified according to the number
of linked antivertices. For example, the one-neighbor vertex is represented
by a vertex-antivertex pair linked to each other by three edges
(see Fig.~\ref{fig:f2}). The concentration of these vertices are denoted
by $\rho_1$, $\rho_2$, or $\rho_3$, respectively, and referred as one-,
two-, and three-neighbor vertices. The total concentration of vertices 
is given as $\rho = \rho_1 + \rho_2 + \rho_3$.

In a previous work \cite{ss:pre02} we have developed a method to determine
the concentration of vertices and also to study some geometrical features
(e.g., arclength measured in lattice constant unit) of the vertex edges.
Now the capacity of this method is extended by allowing a distinction
between the one-, two-, and three-neighbor vertices.
Unfortunately, on a square lattice this methods requires the elimination 
of the four-edge vertices before the geometrical analysis of a given
pattern. It is found, however, that this manipulation causes only a minor
change (much less than one percent) in the Potts (interfacial) energy
except a short transient period. At the same time we can get a more complete
picture of the domain growth. 

\begin{figure}
\centerline{\epsfig{file=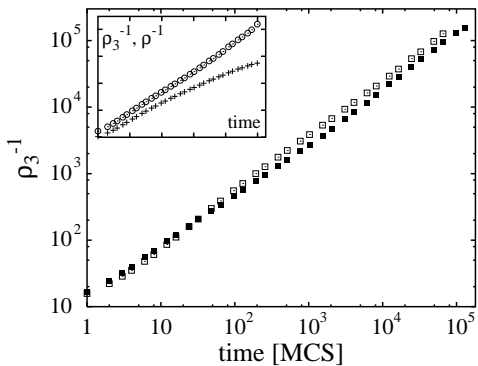,width=6.5cm}} 
\caption{\label{fig:f3} The inverse of $\rho_3$ as a function of time
for the Potts model (filled boxes) and extended voter model (open boxes).
The inset shows the inverse of total concentration of vertices 
$\rho$ (pluses) and the concentration of three-neighbor vertices 
$\rho_3$ (circles) for Potts model. The scales of time
and concentration agree with those of the main plot.}
\end{figure}

Following an earlier suggestion \cite{tainaka:epl91} the inverse of 
vertex concentration can be considered
as a rough estimation of the average area of the growing domains
that increases linearly with time.
The inset of Fig.~\ref{fig:f3} demonstrates that instead of $\rho$,
the concentration of three-neighbor vertices ($\rho_3$) gives a much
better estimate for the expected linear increase in the averaged domain
area for the Potts model. Notice furthermore, that the time dependences
of $\rho_3^{-1}$ are very similar for the Potts and modified voter models
(see Fig.~\ref{fig:f3}) despite the noticeable different behaviors in
$\rho_2$ and $\rho_1$ as discussed below.

\begin{figure}
\centerline{\epsfig{file=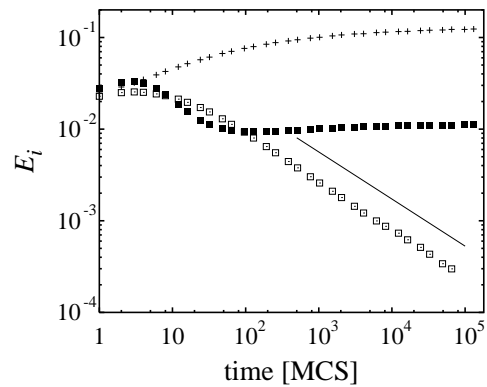,width=6.5cm}}
\caption{\label{fig:f4}The interfacial energies of islands per sites 
as a function of time. The symbols are the same as 
in Fig.~\ref{fig:f1}. The solid line has a slope of $-0.5$.}
\end{figure}

The geometrical analysis allows us to determine the time dependence of
the total perimeter of isolated islands per sites, $E_i$ as a portion of 
interfacial energy $E(t)$. Since we have monitored the average length
of vertex edges ($l_v$) during the simulation,
the total length of vertex edges per sites can be expressed as
$E_v= 3 \rho l_v$. Thus the interfacial energy of islands is 
given as :
\begin{equation}
E_i = E(t) - 3 \rho l_v \; .
\label{eq:li}
\end{equation}
The simulations indicate strikingly different behaviors in $E_i$
for the above three models as plotted  in Fig.~\ref{fig:f4}.
Surprisingly, $E_i$ increases monotonously for the voter model in
the time region in which we could study this system.
It is expected, however, that $E_i$ will decrease for longer times
because it is a part of the total interfacial energy vanishing
as $1 / \ln (t)$ \cite{frachebourg:pre96,bennaim:pre96}.
For the Potts model $E_i$ decreases and tends towards a limit value
dependent on temperature. This limit value, 
which is consistent with the corresponding thermal average value of
Potts energy ($E_T$),
comes from the contribution
of islands generated by the thermal fluctuations within the large domains. 
In the third model $E_i$ approaches asymptotically to an algebraic 
decay ($E_i \propto t^{-1/2}$) manifesting the surface tension-driven
shrink of islands.
Notice that here the dynamical rule
prohibits the creation of islands inside a homogeneous domain.

\begin{figure}
\centerline{\epsfig{file=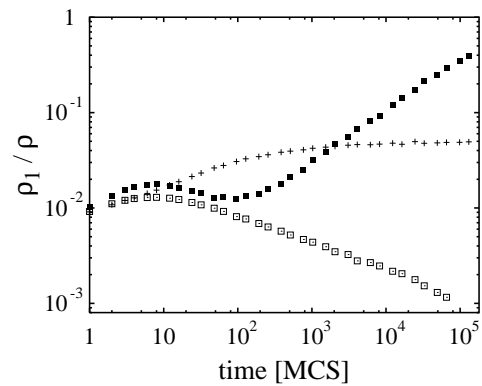,width=6.5cm}}
\caption{\label{fig:f5}The ratio $\rho_1 / \rho$ versus time for the
three models. Symbols as in Fig.~\ref{fig:f1}.}
\end{figure}

The significant differences in the vertex dynamics can also be perceived
when analyzing the $\rho_1 / \rho$ portion of the one-neighbor vertices.
Figure~\ref{fig:f5} shows that $\rho_1 / \rho$ (as well as $\rho_2 / \rho$)
tends towards a fixed ratio for the voter model. It can be assumed that the
ratio $\rho_1 : \rho_2 : \rho_3$ remains fixed for larger times too.
Conversely, in the Potts model the one-neighbor vertices become dominant
for long times because the concentration of the three-neighbor vertices
vanishes as $\rho_3 \propto 1 / t$ (see Fig.~\ref{fig:f3}) meanwhile
$\rho_2 \propto 1 / \sqrt{t}$ in the asymptotic time regime.

For the Potts model the appearance of a new state (e.g., state 0) inside
a domain (of type 1 or 2) represents the birth of a new island. The occurrence 
of this island at the boundaries between the domains of type 1 and 2
yields the creation of a three- or two-neighbor vertex-antivertex
pair (see Fig.~\ref{fig:f2}). In the third model, however,
the extinction of both the one- and two-neighbor vertices as well as of
the islands are driven by interfacial energy (as it happens for the Potts
model).
At the same time their extinction is not compensated by their creation via the
appearance of a new type of domain. 
As a result, the ratio $\rho_1/\rho$ and $\rho_2/\rho$ tends to zero with the
total concentration of vertex in the long time limit.

In the absence of interfacial energy (voter model) the interfaces
become more and more irregular \cite{dornic:prl01} and the
occasional overhanging represents a mechanism to create a new island.
These islands move, change their form, and meet randomly. The meeting of
two islands of the same type represents their fusion and
the contact of two islands of different types creates a three-neighbor
vertex-antivertex pair. Similarly, if an island meets a vertex edge then
a two-neighbor vertex-antivertex pair is created. All of these and the
reversed processes are due to the uncorrelated, random motion of interfaces.
The above results suggest the emergence of some fixed ratio between
the number of one-, two-, and three-neighbor vertices 
in the meantime the typical domain size increases logarithmically.
 
In summary, in the present work we have quantified the topological 
differences occurring during the two-dimensional domain growth 
in three-state systems. The analysis is focused on the time-dependence
of the concentration of the one-, two-, and three-neighbor vertices as
well as on the interfacial energy of islands. The numerical investigations 
indicate that the concentrations of the three types of vertices tend very
slowly towards to fixed ratios while the typical length scale increases
as $r \sim \ln t$ in the voter model. In the Potts model the interfacial
energy results in a faster domain growth (if $T<T_c$) and the vertex
dynamics is governed by the appearance of islands and of the one- and
two-neighbor vertex-antivertex pairs created by the thermal noise.
Above the critical temperature these processes prevent the domain growth.
In the extended voter model the introduction of surface tension changes 
the dynamics of growth dramatically. Similarly to the voter model the
domain growth takes place for arbitrary value of temperature but the
interfacial energy reduces the creation of islands as well as of the 
one- and two-neighbor vertices. As a consequence, in this case
the growth dynamics becomes equivalent to those
characterized by the Allen-Cahn universality class.

\begin{acknowledgments}
This work was supported by the Hungarian National Research Fund under
Grant Nos. F-30449, T-33098 and Bolyai Grant No. BO/0067/00.
\end{acknowledgments}


\end{document}